\documentclass[A4paper]{VisionStyle}
\usepackage{epsfig}

\newcommand{\lnls}{\hbox{Log N--Log S\ }}
\newcommand{\fun}{\hbox{\ erg cm$^{-2}$ s$^{-1}$} }
\newcommand{\lun}{\hbox{\ erg s$^{-1}$} }

\begin{document}

\title{The sources of the X--ray background}

\author{G.\,Hasinger\inst{1,2}} 

\institute{
Max--Planck--Institut f\"ur extraterrestrische Physik, 
Giessenbachstrasse, Garching, D-85741, Germany 
\and 
Astrophysikalisches Institut Potsdam, An der Sternwarte 16, D-14482 
Potsdam, Germany}

\maketitle 

\begin{abstract}

Deep X--ray surveys have shown that the cosmic X--ray background (XRB) is largely
due to the accretion onto supermassive black holes, integrated over 
cosmic time. The {\em ROSAT}, {\em Chandra} and {\em XMM--Newton} satellites have 
resolved more than 80\% of the 0.1--10 keV X--ray background into discrete 
sources. Optical spectroscopic identifications are about 90\% and 60\% complete, 
for the deepest ROSAT and {\em Chandra/XMM--Newton} surveys, respectively, and 
show 
that the sources producing the bulk of the X--ray background are a mixture of 
obscured (type--2) and unobscured (type--1) AGNs, as predicted by the XRB 
population synthesis models, following the unified AGN scenarios. The 
characteristic hard spectrum of the XRB can be explained if most of the AGN are 
heavily absorbed, and in particular a class of highly luminous type--2 AGN, so 
called QSO--2s exist. The deep {\em Chandra} and {\em XMM--Newton} surveys 
have recently 
detected several examples of QSO--2s. The space density of the X--ray selected 
AGN, as determined from ROSAT surveys does not seem to decline as rapidly as 
that of optically selected QSO, however, the statistics of the high--redshift 
samples is still rather poor. The new {\em Chandra} and {\em XMM--Newton} surveys 
at 
significantly fainter fluxes are starting to provide additional constraints 
here, but the preliminary observed redshift distribution peaks at much lower 
redshifts (z=0.5--0.7) than the predictions based on the ROSAT luminosity function.

\keywords{Missions: XMM--Newton, Chandra -- Subjects: Cosmology, Backgrounds, AGN \ }
\end{abstract}

\section{Introduction}

\begin{figure*}[!ht]
\begin{center}
\epsfig{file=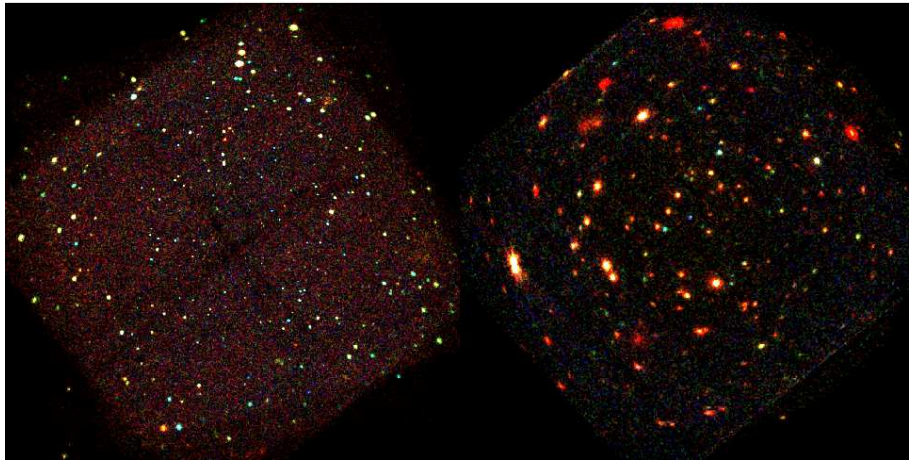, width=12cm}
\end{center}
\caption[]{ {\em Chandra} ACIS--I image of the {\em Chandra Deep Field South} (left, \cite{ghasinger-F1:Ros02})
and {\em XMM--Newton} EPIC image of the {\em Lockman Hole} field (right, \cite{ghasinger-F1:Has01a}).  
The field sizes are about 20 x 20 arcmin and 30 x 30 arcmin, respectively. The colours refer to 
photons detected  in different energy bands: red, green 
and blue correspond to the soft, medium and hard X--ray range, 
respectively.}
\label{ghasinger-F1:Fig1}
\end{figure*}

Deep X--ray surveys indicate that the cosmic X--ray background (XRB) is largely 
due to accretion onto supermassive black holes, integrated over cosmic time. In 
the soft (0.5--2 keV) band more than 90\% of the XRB flux has been resolved using a 
1.4 Msec observation with {\em ROSAT} (\cite{ghasinger-F1:Has98}) and recently two 1 Msec {\em
Chandra} 
observations (\cite{ghasinger-F1:Ros02,ghasinger-F1:Bra01b}) and a 100 ksec observation with {\em XMM--Newton} 
(\cite{ghasinger-F1:Has01a}) (see Figure 
\ref{ghasinger-F1:Fig1}). In the harder (2--10 keV) band a similar fraction of 
the background has been resolved with the above {\em Chandra} and 
{\em XMM--Newton surveys}, 
reaching source densities of about 4000 deg$^{-2}$. Surveys in the very hard 
(5--10 keV) band have been pioneered using {\em BeppoSAX}, which resolved about 
30\% of 
the XRB (\cite{ghasinger-F1:Fio99}). {\em XMM--Newton} and {\em Chandra} have now also 
resolved the majority 
(60--70\%) of the very hard X--ray background. The \lnls distribution shows a 
significant cosmological flattening in the softer bands (see Figure 
\ref{ghasinger-F1:Fig2}), while in 
the very hard band it is still relatively steep, indicating that those surveys 
have not yet sampled the redshifts where the strong cosmological evolution of 
the sources saturates.

Optical followup programs with 8--10m telescopes have been completed for the 
ROSAT deep surveys and find predominantly Active Galactic Nuclei (AGN) as 
counterparts of the faint X--ray source population (\cite{ghasinger-F1:Schm98,ghasinger-F1:Zam99,ghasinger-F1:Leh01}) 
mainly X--ray and optically unobscured AGN (type--1 Seyferts and QSOs) and a 
smaller fraction of obscured AGN (type--2 Seyferts). Optical identifications for 
the deepest {\em Chandra} and {\em XMM--Newton} fields are still far from
complete, however a 
mixture of obscured and unobscured AGN with an increasing fraction of 
obscuration seems to be the dominant population in these samples, too 
(\cite{ghasinger-F1:Fio00,ghasinger-F1:Bar01a,ghasinger-F1:Toz01,ghasinger-F1:Ros02}; 
see below). Interestingly, first examples of 
the long--sought class of high--redshift, high--luminosity, heavily obscured 
active 
galactic nuclei (type--2 QSO) have been detected in deep {\em Chandra} fields 
(\cite{ghasinger-F1:Nor01,ghasinger-F1:Ste01}) and in the {\em XMM--Newton} Deep survey in the {\em Lockman
Hole} field (\cite{ghasinger-F1:Has01b}).

After having understood the basic contributions to the X--ray background, the 
general interest is now focussing on understanding the physical nature of these 
sources, the cosmological evolution of their properties, and their role in 
models of galaxy evolution. We know that basically every galaxy with a 
spheroidal component in the local universe has a supermassive black hole in its 
center (\cite{ghasinger-F1:Ge00}). The luminosity function of X--ray selected AGN shows strong 
cosmological density evolution at redshifts up to 2, which goes hand in hand 
with the cosmic star formation history (\cite{ghasinger-F1:Miy00,ghasinger-F1:Miy01}). At the redshift peak 
of optically selected QSOs around z=2.5 the AGN space density is several hundred 
times higher than locally, which is in line with the assumption that most 
galaxies have been active in the past and that the feeding of their black holes 
is reflected in the X--ray background. While the comoving space density of 
optically and radio--selected QSO declines significantly beyond a redshift of 
3 (\cite{ghasinger-F1:Ssg95,ghasinger-F1:Fan01,ghasinger-F1:Sha96}), a similar decline has not yet been observed in the 
X--ray selected AGN population (\cite{ghasinger-F1:Miy00}), although the statistical quality of 
the 
high--redshift AGN samples needs to be improved. The new {\em Chandra} and 
{\em XMM--Newton} surveys are bound to give additional constraints here.    

The X--ray observations have so far been about consistent with population 
synthesis 
models based on unified AGN schemes (\cite{ghasinger-F1:Com95,ghasinger-F1:Gil01}), which explain the hard 
spectrum of the X--ray background by a mixture of absorbed and unabsorbed AGN, 
folded with the corresponding luminosity function and its cosmological 
evolution. According to these models, most AGN spectra are heavily absorbed and 
about 80\% of the light produced by accretion will be absorbed by gas and dust 
(\cite{ghasinger-F1:Fab98}). However, these models are far from unique and contain a number of 
hidden assumptions, so that their predictive power remains limited until 
complete samples of spectroscopically classified hard X--ray sources are 
available. In particular they require a substantial contribution of 
high--luminosity obscured X--ray sources (type--2 QSOs), which so far have only 
scarcely 
been detected. The cosmic history of obscuration and its potential dependence 
on 
intrinsic source luminosity remain completely unknown. 
\cite*{ghasinger-F1:Gil01} e.g. assumed strong evolution of the obscuration fraction 
(ratio of type--2/type--1 AGN) from 4:1 in the local universe to much larger 
covering fractions (10:1) at high redshifts (see also \cite{ghasinger-F1:Fab98}). The gas to 
dust ratio in high--redshift, high--luminosity AGN could be completely different 
from the usually assumed galactic value due to sputtering of the dust particles 
in the strong radiation field (\cite{ghasinger-F1:Gra97}). This might provide objects which are
heavily absorbed at X--rays and unobscured at optical wavelengths.

In this paper I shortly discuss the current status of the optical identification 
work in the {\em ROSAT/XMM--Newton}/ {\em Chandra} deep survey in the {\em Lockman 
Hole}, which is largely 
based on optical work with the Keck telescope led by Maarten Schmidt (see 
\cite{ghasinger-F1:Schm98} and \cite{ghasinger-F1:Has01b} for more detail). I also present preliminary 
results of optical identifications in the {\em Chandra Deep Field South},
obtained 
with the ESO VLT, which will be formally published in \cite{ghasinger-F1:Szo02} (see also 
\cite{ghasinger-F1:Toz01,ghasinger-F1:Ros02}). I then 
discuss the results and try to come to some tentative conclusions about the 
evolution of X--ray sources at high redshifts.

\section{Optical identifications of deep X--ray surveys}

\subsection{The {\em Lockman Hole} field}

The {\em Lockman Hole} field has been observed with the {\em XMM--Newton} observatory
during the performance verification phase (see Figure \ref{ghasinger-F1:Fig1}a). About 100 ksec good
data, centered on the same sky position as the {\em ROSAT HRI} pointing, have
been accumulated with the European Photon Imaging Camera (EPIC) reaching
minimum fluxes of 0.31, 1.4 and 2.4 $\cdot$ 10$^{-15}$ erg cm$^{-2}$ s$^{-1}$
in the 0.5--2, 2--10 and 5--10 keV energy bands. Within an off--axis angle of 10 
arcmin 148, 112 and 61 sources, respectively, have been detected. In the 5--10
keV energy band a somewhat lower sensitivity compared to the 1Msec 
{\em Chandra Deep Field South} observation has been reached (see \cite{ghasinger-F1:Ros02}),
resolving $\sim$60 \% of 
the very hard X--ray background (\cite{ghasinger-F1:Has01a}). This is about a factor of 20 more 
sensitive than the previous {\em BeppoSAX} observations. A total of 300 ksec 
observations of the Lockman Hole has been accumulated with the {\em Chandra} HRC 
in the 0.5--7.0 keV band, reaching a similar flux limit compared to the {\em 
XMM--Newton} pointing (\cite{ghasinger-F1:Mur02}). The {\em Chandra} HRC data provide very 
accurate 
source positions, whereas {\em XMM--Newton} allows spectrophotometry of very 
faint intrinsically absorbed X--ray sources due to its unprecedented sensitivity
in the hard band.

\begin{figure}[!ht]
\begin{center}
\epsfig{file=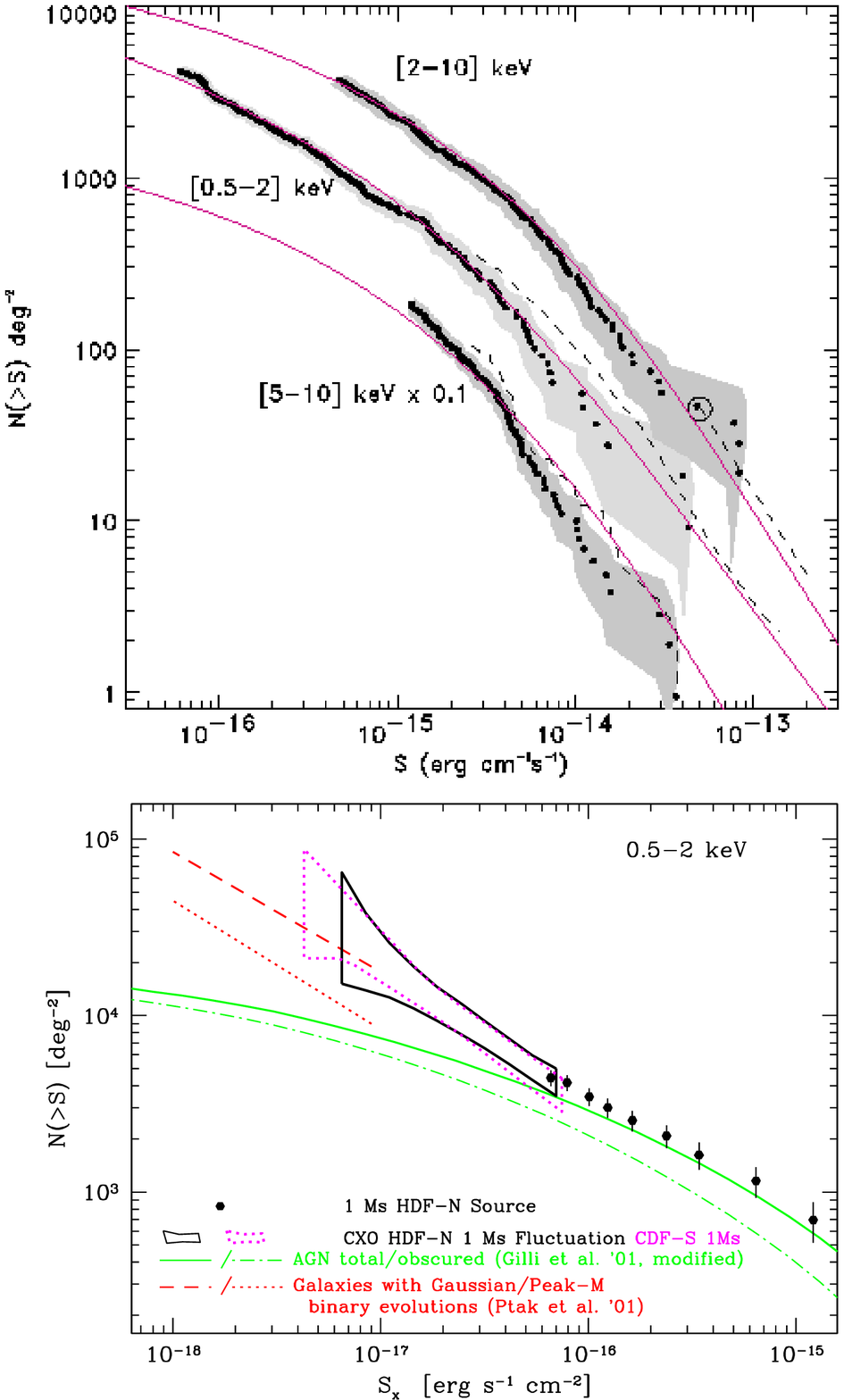, width=8cm}
\end{center}
\caption[]{Top: \lnls in the soft (0.5--2.0 keV), hard (2--10 keV) and
very hard (5--10 keV) energy band from the 940 ksec {\em Chandra 
Deep Field South} observation (\cite{ghasinger-F1:Ros02}). The dashed 
lines in the soft and very hard band are source counts from the 
{\em Lockman Hole} field from {\em ROSAT} (\cite{ghasinger-F1:Has93}) and 
{\em XMM-Newton} (\cite{ghasinger-F1:Has01a}), respectively; in the hard
band they refer to the {\em ASCA} survey by \cite*{ghasinger-F1:Cag98}. 
The thin solid lines give the source counts predicted by the most 
recent background synthesis models (\cite{ghasinger-F1:Gil01}).
Bottom: fluctuation analysis from the HDF--N and CDFS megasecond
Chandra 
observations from \cite*{ghasinger-F1:Miy02}. At the faintest fluxes
there is a significant source count excess above the predictions of the
\cite*{ghasinger-F1:Gil01} AGN background population synthesis model
(solid line), which is most likely due to normal galaxies.} 
\label{ghasinger-F1:Fig2}
\end{figure}

The optical counterparts for $\sim$60 X--ray sources are already known from
the spectroscopic identification of the {\em ROSAT Ultradeep Survey} sample 
(\cite{ghasinger-F1:Leh01}). Among them are 
one of the most distant X--ray selected quasars at $z=4.45$ (\cite{ghasinger-F1:Schn98}) and 
one of the highest redshift, probably merging cluster of galaxies at $z=1.26$ 
(\cite{ghasinger-F1:Has99,ghasinger-F1:Tho01,ghasinger-F1:Hash02}). We have identified 25 new 
{\em XMM--Newton} sources using low--resolution multi--slit mask spectra taken with the 
LRIS instrument at the Keck II telescope in March 2001 (PI: M. Schmidt; 
\cite{ghasinger-F1:Leh02}). 
Among the new XMM sources we have found only a few new broad emission line AGNs 
(type--1), while the optical spectra of most new sources show narrow emission 
lines and/or only galaxy--like continuum emission at redshifts $z<1.0$. In 
several cases high ionisation emission lines like $[$Ne~V$]$ $\lambda3426$ are 
absent and thus we see no sign for AGN activity in the optical spectrum, 
however, their high X--ray luminosity (L$_{X}>10^{43}$ erg s$^{-1}$) and/or the 
strong intrinsic absorption (log N$_{H}>22.0~cm^{-2}$) reveal a type--2 AGN in 
these sources. Three new sources showing typical galaxy spectra, have been 
detected only in the 0.5--2.0 keV band. Due to their relatively low X--ray 
luminosities (log L$_{X}<42.0$) and their soft X--ray spectra (no indication for 
intrinsic absorption) we classify them as normal galaxies. Several sources with 
X--ray luminosities in the range of $42.0<$ log L$_{X}<43.0$, which show galaxy--like 
optical spectra, are hard to classify due to the small number of photons in 
their X--ray spectra. We preliminarily classify them as type--2 AGN/galaxy. The 
completeness of the identification ranges from 61\% in the soft sample to 79\% 
in the ultra hard sample (5--10 keV energy band) (\cite{ghasinger-F1:Leh02}). The majority of 
the so far spectroscopically identified sources are type--1 and type--2 AGNs. 
Although we have no complete identification so far we find a strong indication 
for a larger fraction of type--2 AGNs, especially in the ultra hard sample, 
compared to that ($\sim$20\%) of the UDS. Nearly all spectroscopically 
identified type--2 AGNs are at moderate redshift ($z<1$). One type--2 QSO 
candidate (X174A) at $z=3.240$ has been identified in the Lockman Hole region 
so far (\cite{ghasinger-F1:Has01b}). 

Most of the unidentified faint {\em 
XMM--Newton} sources have very faint optical counterparts ($R>24.0$) and at 
least half of them are extremely red objects (EROs, $R-K'>5.0$,
see also \cite{ghasinger-F1:Ber02}).
The new {\em XMM--Newton} sources with EROs as optical counterparts are similar 
to those objects in the UDS with photometric redshifts suggesting obscured AGNs 
at redshifts $1<z<3$. The photometric redshift technique is probably the
only tool to identify such faint optical objects.
The {\em XMM--Newton} source population at faint fluxes is therefore likely 
dominated by obscured AGNs (type--2), as predicted by the AGN population 
synthesis models for the X--ray background.

\begin{figure}[!ht]
\begin{center}
\epsfig{file=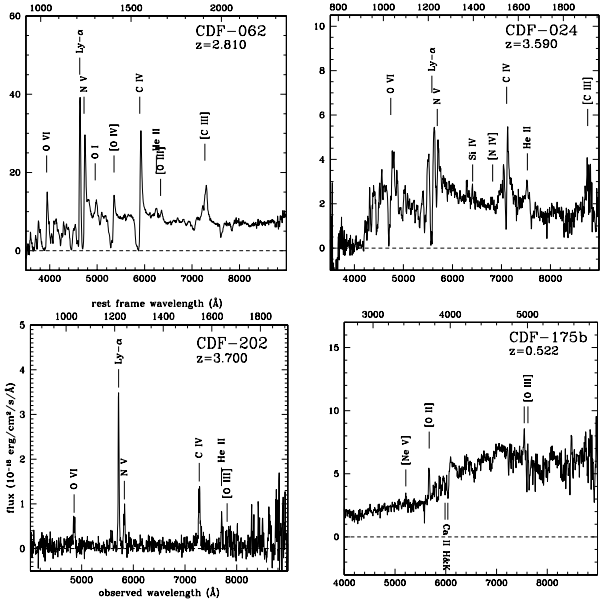, width=8cm}
\end{center}
\caption[]{Optical spectra of some selected CDFS sources obtained using 
multiobject--spectroscopy with FORS at the VLT (\cite{ghasinger-F1:Szo02}). a: broad absorption 
line (BAL) QSO CDFS--062 at z=2.822 (see also \cite{ghasinger-F1:Gia01}); b: high--redshift QSO 
CDFS--024 at z=3.605, showing strong absorption lines; c: QSO--2 CDFS--202 at 
z=3.700 (see \cite{ghasinger-F1:Nor01}); d: Seyfert--2 CDFS--175b at z=0.522, showing a weak
high excitation line of [NeV].}
\label{ghasinger-F1:Fig3}
\end{figure}

\subsection{The {\em Chandra Deep Field South} (CDFS)}

The {\it Chandra X--ray Observatory} has performed deep X--ray surveys 
in a number of fields with ever increasing exposure 
times (\cite{ghasinger-F1:Mus00,ghasinger-F1:Hor00,ghasinger-F1:Gia01,ghasinger-F1:Toz01,ghasinger-F1:Bra01a}) and has recently completed two 
1~Megasec {\it Chandra} exposures, in the {\em Chandra Deep Field South} (CDFS, 
\cite{ghasinger-F1:Gia02,ghasinger-F1:Ros02}) and in the {\em Hubble Deep Field North} (HDF--N,
\cite{ghasinger-F1:Bra01b}), the latter exposure is currently being increased to 2~Megasec.

Here I discuss results from the 940 ksec CDFS observation. 
The source counts (see Figure \ref{ghasinger-F1:Fig2}) have been extended to $5.5 \times 10^{-17}\fun$ in
the soft 0.5--2 keV band and $4.5 \times 10^{-16}\fun $ in the hard
2--10 keV band, reaching a space density of almost $4000~deg^{-2}$,
resolving $> 80\%$ of the background in both bands. A total of 346 sources
has been detected (\cite{ghasinger-F1:Toz01,ghasinger-F1:Ros02}). 

Deep optical imaging and multiobject spectroscopy has been performed in 12 nights with the 
ESO Very Large Telescope (VLT) in the time frame April 2000 -- December 2001,
using the FORS 
instruments with individual exposure times ranging from 1--5 hours.
Some preliminary results including the VLT optical spectroscopy have already 
been presented elsewhere (\cite{ghasinger-F1:Nor01,ghasinger-F1:Toz01,ghasinger-F1:Ros02}). 
A complete optical spectroscopy catalogue 
will be published by \cite*{ghasinger-F1:Szo02}. Figure 
\ref{ghasinger-F1:Fig3} shows examples of 4 VLT spectra.
The upper two spectra show high--redshift QSOs with restframe--UV absorption 
features (BAL or mini--BAL QSOs), which both have some indication of intrinsic 
absorption in their X--ray spectra. The object in the lower left is the famous,
highest redshift type--2 QSO detected in the CDFS with heavy X--ray absorption in
the QSO rest frame (\cite{ghasinger-F1:Nor01}). The spectrum in the lower right shows a 
Seyfert--2 galaxy with heavy X--ray absorption and an AGN--type luminosity. The
latter spectrum is characteristic for the bulk of the detected galaxies, which 
show either no or very faint high excitation lines indicating the AGN nature of 
the object, so that we have to resort to a combination of optical and X--ray 
diagnostics to classify them as AGN (see below). Redshifts could be obtained
so far for 169 of the 360 sources in the CDFS, of which 123 are very reliable
(high quality spectra with 2 or more spectral features), while the remaining
optical spectra contain only a single emission line, or are too noisy.
For objects fainter than R=24 reliable redshifts can be obtained (see also
Figure \ref{ghasinger-F1:Fig5}), if the spectra
contain strong emission lines. For the remaining optically faint objects we 
have to resort to photometric redshift techniques. About 11\% of the CDFS 
sources have 
no counterpart even in deep VLT optical images ($R<27.5$) or near--IR imaging
(15\% at $K<22$) (\cite{ghasinger-F1:Ros02}). Nevertheless, for a subsection of the sample at
off--axis angles smaller than 8 arcmin we obtain a spectroscopic completeness of 
about 60\%.  

\begin{figure}[!ht]
\begin{center}
\epsfig{file=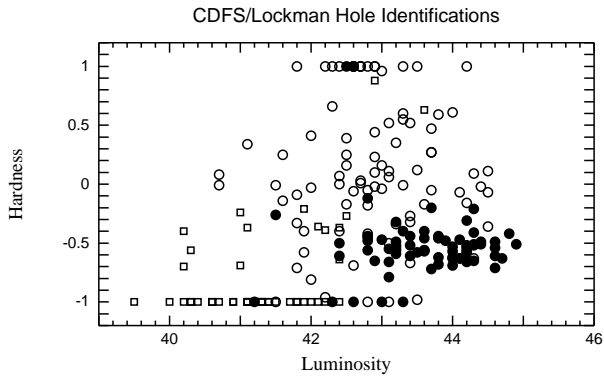, width=8cm}
\end{center}
\caption[]{Hardness ratio of CDFS and Lockman Hole Sources with redshifts
as a function of X--ray luminosity. 
Filled circles correspond to type--1 AGN and are clustered in a narrow band of 
relatively soft hardness ratios. Open circles correspond to type--2 AGN, either 
classified through their optical spectra as Seyfert--2 galaxies, or through
the combination of hard hardness ratio and high X-ray luminosity. Open squares
correspond to normal galaxies, classified by their typically very soft X--ray spectrum
and low X--ray luminosity. In particular in the range of soft hardness ratios
($-0.7 , HR , -0.3$) and low X--ray luminosities ($log(LX) < 42.5$) there are 
ambiguities in this classification.}
\label{ghasinger-F1:Fig4}
\end{figure}

\subsection{Optical/X--ray classification}

Type--1 AGN (Seyfert--1 and QSOs) can be often readily identified by 
the broad permitted emission lines in their optical spectra.
Luminous Seyfert--2 galaxies show strong forbidden emission lines
and high--excitation lines indicating photoionization by a hard continuum 
source. 
However, already in the spectroscopic identifications of the 
{\em ROSAT} Deep Surveys it became apparent, that an increasing 
fraction of faint X--ray selected AGN shows a significant, sometimes
dominant contribution of stellar light from the host galaxy in
their optical spectra, depending on the ratio of optical luminosity
between nuclear and galaxy light (\cite{ghasinger-F1:Leh00,ghasinger-F1:Leh01}). If an AGN
is outshined by its host galaxy it is not possible to detect it 
optically. Many of the 
counterparts of the faint X--ray sources detected by {\em Chandra} 
and {\em XMM--Newton} show optical spectra which are dominated
by their host galaxy and only a minority has clear indications of
an AGN nature (see also \cite{ghasinger-F1:Bar01a,ghasinger-F1:Bar01b}). In these cases, the
X--ray emission could still be dominated by the active galactic 
nucleus, while a contribution from stellar and thermal processes
(hot gas from supernova remnants, starbursts and thermal halos, or
a population of X--ray binaries) can be important as well.
     
In these cases X--ray diagnostics in addition to the optical
spectroscopy can be crucial to classify the source of the 
X--ray emission. AGN have typically (but not always!) X--ray 
luminosities above $10^{42}\lun$ and power law spectra, often 
with significant intrinsic absorption. Local, well--studied 
starburst galaxies have integrated 
X--ray luminosities typically below
$10^{42}\lun$ and very soft X--ray spectra. Thermal haloes
of galaxies and the intergalactic gas in groups can have
higher X--ray luminosities, but has soft spectra as well.
The redshift effect in addition helps the X--ray diagnostic,
because soft X--ray spectra appear even softer already at
moderate redshift, while the typical AGN power law spectra 
appear harder over a very wide range of redshifts. 

In Figure 
\ref{ghasinger-F1:Fig4} the X--ray hardness ratio is shown as a function of the
X--ray luminosity (in the 0.5--2 keV, 2--10 keV, or 0.5--10 keV 
band, depending on in which band the object was detected) for 
170 sources for which we have optical spectra and reliable redshifts
in the CDFS (\cite{ghasinger-F1:Szo02}; see also \cite{ghasinger-F1:Ros02}) and the 
{\em Lockman Hole} (\cite{ghasinger-F1:Leh01,ghasinger-F1:Leh02})
for X--ray sources detected by {\em Chandra} and {\em XMM--Newton},
respectively. The hardness ratio is defined as $HR \equiv
(H-S)/(H+S)$ where $H$ and $S$ are the net count rates in the hard
(2--7 keV for {\em Chandra} and 2--4.5 keV for {\em XMM--Newton}) and the soft band 
(0.5--2 keV), respectively.  The X--ray
luminosities are not corrected for internal absorption and are
computed assuming $H_0=50$ km s$^{-1}$ Mpc$^{-1}$ and $q_0=0.5$.

Although this diagram is for illustration purposes only and a 
a correct treatment would have to properly take into account 
the different instrument characteristics and detection bands,
it clearly shows a segregation of the different X--ray emitters
(indicated by the dashed elliptical outlines. Type--1
AGNs have luminosities above $10^{42}\lun$ and hardness ratios 
scattered around $HR\simeq -0.6$, corresponding to a power law 
photon index around $\Gamma = 1.9$, typical for the intrinsic 
continuum of AGN. Type--2 AGN have observed luminosities above 
$10^{40.5}\lun$ (intrinsically higher), but are scattered to much larger hardness
ratios ($HR>-0.2$). Direct spectral fits of the XMM--Newton and 
(some) Chandra spectra clearly indicate, that these harder spectra
are due to neutral gas absorption and not due to a flatter intrinsic
slope (\cite{ghasinger-F1:Nor01,ghasinger-F1:Has01b,ghasinger-F1:Ber02,ghasinger-F1:Mai02}). 
It is interesting to note that
no high--luminosity, very hard sources exist in this diagram.
This is due to a selection effect of the pencil beam surveys:
due to the small solid angle, the rare high luminosity sources
are only sampled at high redshifts, where the absorption
cutoff of type--2 AGN is redshifted to softer X--ray energies.
Indeed, the type--2 QSOs in this sample (\cite{ghasinger-F1:Nor01,ghasinger-F1:Leh02})
are the objects at $L_X> 10^{44}\lun$ and $HR > 0$.

About 10\% of the sources have
optical spectra of normal galaxies, X--ray luminosities below 
$10^{42}\lun$ and very soft spectra ($HR \sim-1$), typical
for starburst galaxies or hot gas halos. The deep 
{\em Chandra} and {\em XMM--Newton} surveys therefore for 
the first time detect the population of normal starburst 
galaxies out to intermediate redshifts 
(\cite{ghasinger-F1:Mus00,ghasinger-F1:Gia01,ghasinger-F1:Leh02}), for which a 
significant contribution to the XRB had been claimed
for a long time (e.g. \cite{ghasinger-F1:McH98}). Those galaxies, however, 
appear at much lower fluxes
and therefore produce an almost negligible contribution to the 
background. They might become an important means
to study the star formation history in the universe completely
independently from optical/UV, sub--mm or radio observations.
However, in the X--ray luminosity range around $10^{42}\lun$,
where the emission from star forming processes and the 
central AGN may be comparable, there will always remain
ambiguities.

The joint optical/X--ray diagnostics scheme can also be 
applied to the spectroscopically identified X--ray 
sources in other deep {\em Chandra} fields
in order to obtain an as complete sample of faint 
X--ray source classification as possible. Table 1 gives 
a summary of optical identifications and X--ray source types
in the two deep fields discussed here, as well as in 
the {\em Hawaii 13hr} field, the {\em Abell 370} cluster field 
and the {\em Hubble Deep Field North} which all have 
spectroscopically identified samples in the 
literature (\cite{ghasinger-F1:Bar01a,ghasinger-F1:Bar01b}).

\begin{table}
\caption[]{{\it Chandra} and {\it XMM--Newton} Survey Identifications}
\begin{center}
\begin{tabular}{lrrrl}
\hline\hline
Field   &        AGN1 & AGN2 & Gal. & Reference \\
\hline
LHole  &  41     &  26    &  7       &\cite{ghasinger-F1:Leh02} \\
CDFS          &  47     &  73    & 49       &\cite{ghasinger-F1:Szo02} \\
Abell370     &   9     &   5    &  6        &\cite{ghasinger-F1:Bar01b} \\
13hr$^a$      &   5     &   7    &  1       &\cite{ghasinger-F1:Bar01a} \\
HDF--N$^a$     &  10     &  10    &  0      &\cite{ghasinger-F1:Bar01b} \\
\hline
\end{tabular}
\label{ghasinger-F1:tab_surv}
\end{center}
\centerline{$^{\rm a}$  only 2--7 keV band detections considered}
\end{table}

\begin{figure}[!ht]
\begin{center}
\epsfig{file=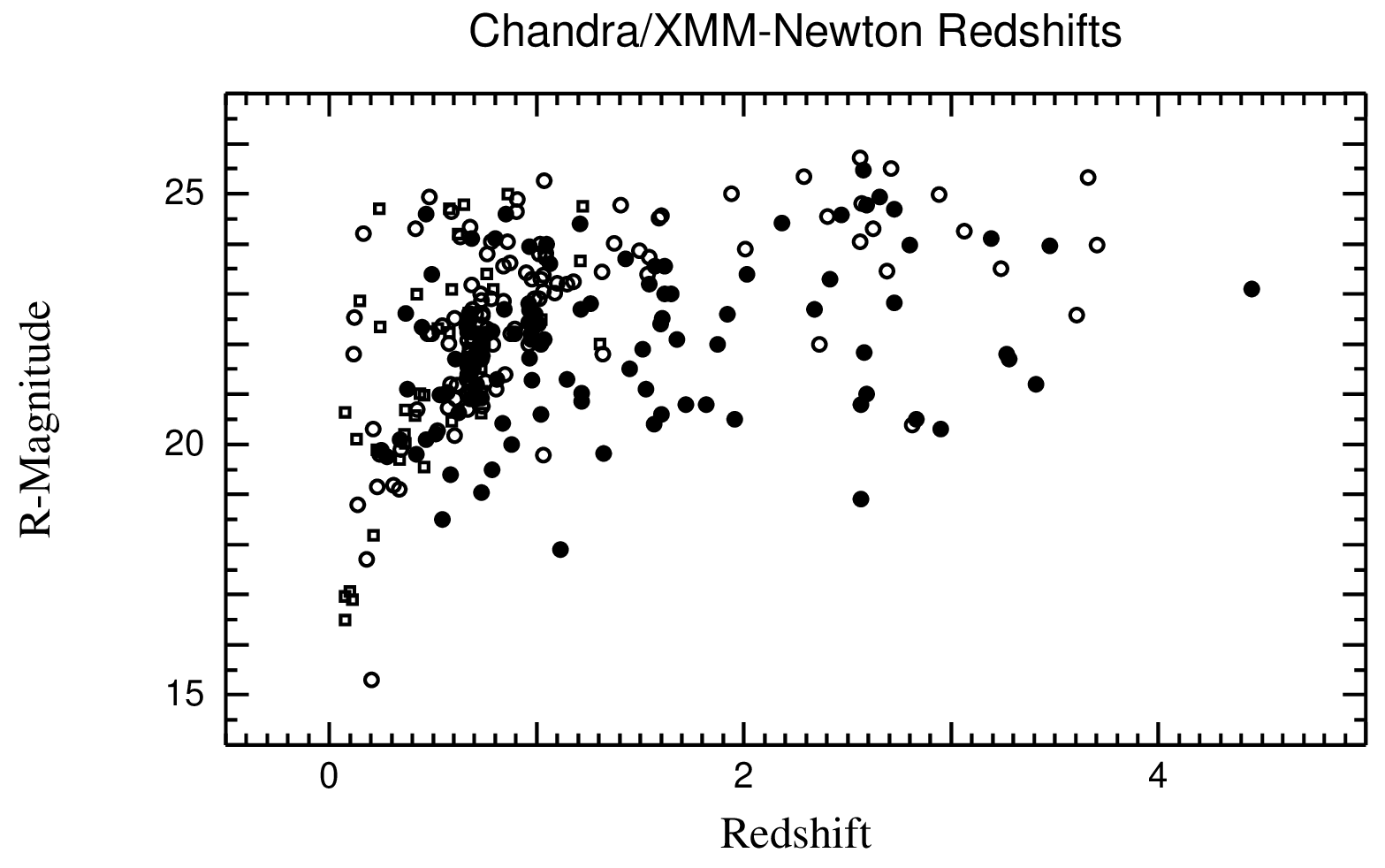, width=8cm}
\epsfig{file=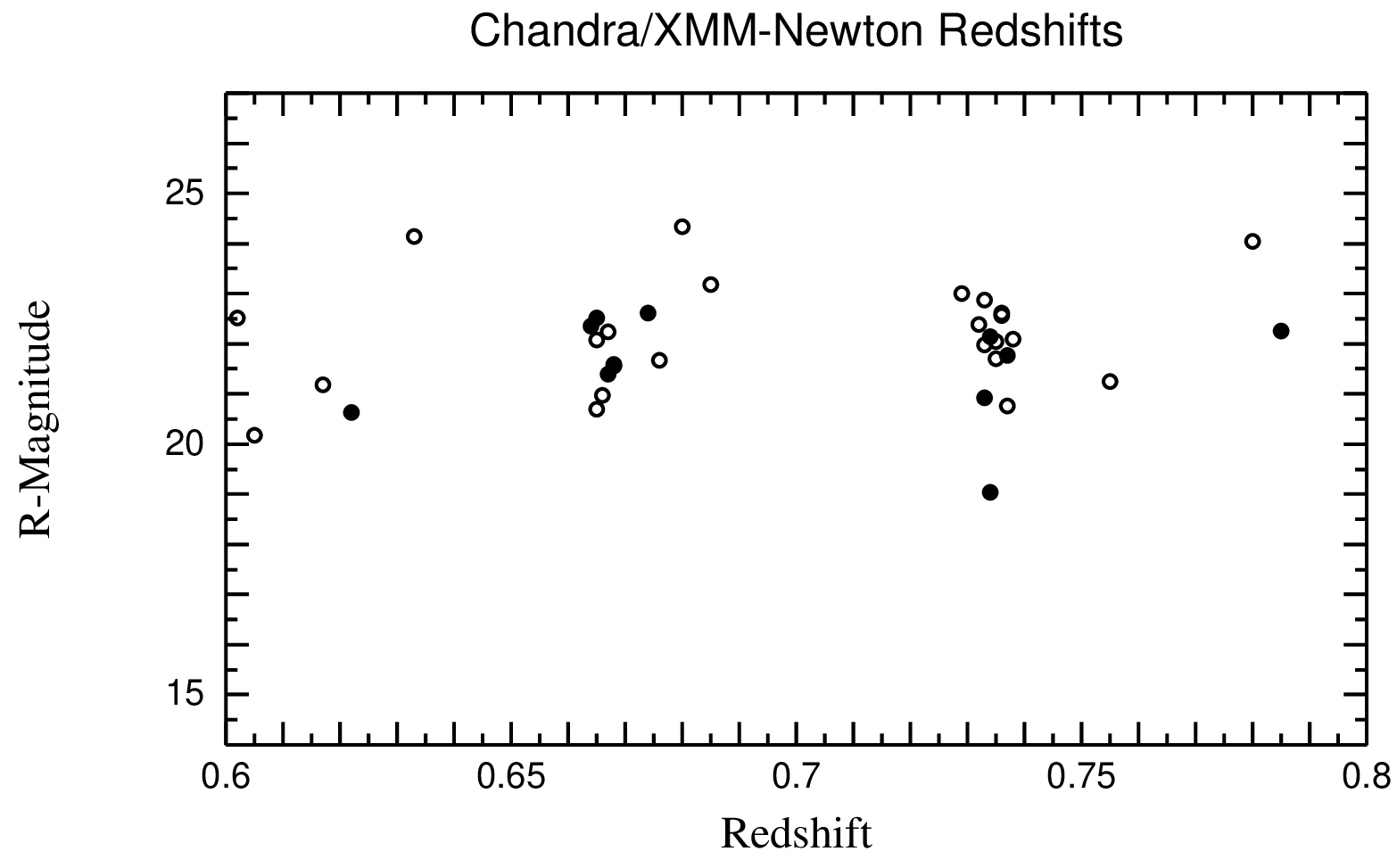, width=8cm}
\end{center}
\caption[]{Top: Optical magnitudes of AGN and galaxies from
all the samples in table 1 as a function of redshift. R--magnitudes
are taken from the Lockman Hole (\cite{ghasinger-F1:Leh01,ghasinger-F1:Leh02}) 
and the CDFS 
samples (\cite{ghasinger-F1:Gia02}). For the Hawaii 13hr (\cite{ghasinger-F1:Bar01a}), 
Abell 370, and HDF--N samples (\cite{ghasinger-F1:Bar01b}), where only I magnitudes are
given, a colour of R--I=1 has been assumed. Bottom: the CDFS sample in the
redshift range 0.6-0.8. An accumulation of objects
in two redshift bins around z=0.7 is due to large scale structure
in the CDFS.}
\label{ghasinger-F1:Fig5}
\end{figure}

\section{The new redshift distribution}

\begin{figure}[!ht]
\begin{center}
\epsfig{file=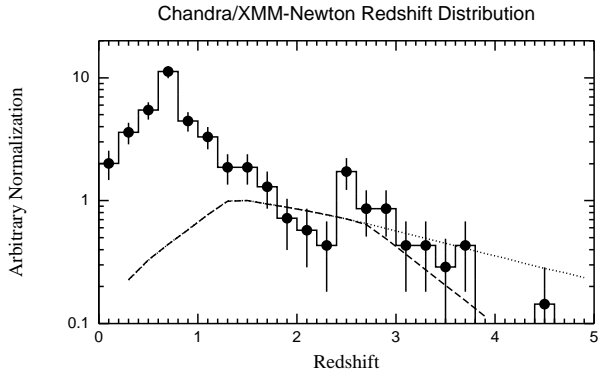, width=8cm}
\end{center}
\caption[]{Redshift distribution of $\sim 300$ X--ray selected
AGN and galaxies in the deep Chandra and XMM--Newton survey samples
given in table 1 (solid circles and histogram), compared to
model predictions from population synthesis models (\cite{ghasinger-F1:Gil01}).
The dashed line shows the prediction for a model, where 
the commoving space density of high--redshift QSO follows the
decline above z=2.7 observed in optical samples (\cite{ghasinger-F1:Ssg95,ghasinger-F1:Fan01}). 
The dotted line shows a prediction with a constant space density 
for $z>1.5$. The two model curves have been normalized to their
peak at z=1, while the observed distribution has been normalized 
to roughly fit the models in the redshift range 1.7--3.}
\label{ghasinger-F1:Fig6}
\end{figure}

All the above samples have a spectroscopic completeness of about 
60\%, which is mainly caused by the fact that about 40\% of the 
counterparts are optically too faint to obtain reliable spectra. 
This incompleteness is probably also reflecting some redshift
bias, most likely higher redshift objects are missing, as well
as faint emission line objects, where the strongest emission 
lines ([OII], Ly$_\alpha$) fall outside the optical bands. On
the other hand, the optically faintest identified sources 
(R=24--25) are distributed throughout the whole redshift range
z=0--4 (see Figure \ref{ghasinger-F1:Fig5}), therefore there is reason to believe that 
a substantial
fraction of the so far unidentified sources follows the same
redshift distribution as the identified sources. The completeness 
of 60\% therefore allows to compare the redshift distribution
with predictions from X--ray background population synthesis 
models (\cite{ghasinger-F1:Gil01}), based on the AGN X--ray luminosity function and
its evolution as determined from the ROSAT surveys (\cite{ghasinger-F1:Miy00}),
which predict a maximum at redshifts around z=1.5. 
Figure \ref{ghasinger-F1:Fig6}
shows two predictions of the redshift distribution from the 
\cite*{ghasinger-F1:Gil01} model for a flux limit of 
$2.3 \times 10^{-16}~\fun$ in the 0.5--2 keV band with 
different assumptions for the high--redshift evolution of
the QSO space density. The two models from \cite*{ghasinger-F1:Gil01} 
have been normalized at the peak of the distribution.

The actually observed redshift distribution does not vary
significantly within the flux limit range covered by the 
samples in table 1, therefore the total observed redshift distribution
is shown in Figure \ref{ghasinger-F1:Fig6} for the total number of $\sim 300$
sources in all samples. The observed redshift distribution,
arbitrarily normalized to roughly fit the population synthesis
models in the redshift range 1.5 -- 3 keV is radically different 
from the prediction, with a peak at a redshift in the 
range 0.5--0.7. This is still the case, if the objects belonging
to the large scale structures around z=0.7 in the CDFS 
are removed. The total number of objects at redshift less
than 1 is significantly higher than the model predictions,
even ignoring the 40\% spectroscopic incompleteness. The 
peak at redshifts below 1 is also significant, if the 
normal star forming galaxies in the sample are removed.  
This clearly demonstrates that the population synthesis 
models will have to be modified to incorporate different
luminosity functions and evolutionary scenarios
for intermediate--redshift, low--luminosity AGN.

\section{The AGN evolution at high redshift}

The comparison between the observed and predicted N(z) distributions
at high redshifts is complicated by the possible existence of  
large--scale structure in the pencil beam survey (there is e.g. a
possibly significant excess of objects around z=2.5 in the CDFS),
but also by redshift--dependent selection effects and in general by
the still relatively small volume sampled and therefore poor 
counting statistics in the number of objects. In addition, the 
overall normalization of the curves is uncertain because of the 
significant mismatch of the distribution at low z. Nevertheless, 
the observed distribution is roughly consistent with both predictions
in the redshift range z=1.6--3.8. There is, however, a significant 
discrepancy between the observed distribution and the constant 
space density model (dotted line) at redshifts above 4, where only 
one object was detected, while about 8 objects would be predicted. 
From Figure \ref{ghasinger-F1:Fig5} it becomes apparent, that the dearth of 
X--ray selected 
AGN is probably not due to optical spectroscopic selection effects.
The one object detected at z=4.45 already in the ROSAT data of the 
Lockman Hole (\cite{ghasinger-F1:Schn98}) has an optical magnitude of R=23 and is
therefore not at the spectroscopic limit of the samples. Also the
$Ly_\alpha$ and CIV lines for QSOs in the redshift range 4--5 fall
well into the optical range. The observed redshift distribution
therefore gives a strong indication for a decline of the QSO 
space density beyond a redshift of 3.8.     

\begin{figure}[!ht]
\begin{center}
\epsfig{file=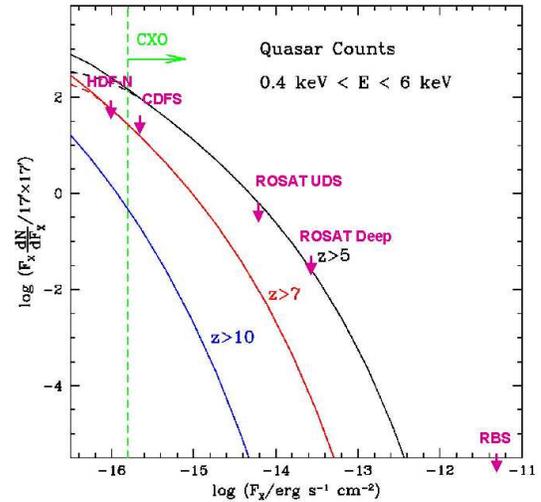, width=7cm}
\end{center}
\caption[]{Prediction of the number density of AGN with redshits larger 
than 5, 7 and 10, respectively as a function of flux in a typical $17 \times 17$
arcmin Chandra field of view from \cite*{ghasinger-F1:Hai99}. 
Upper limits measured in X--ray surveys at 
various flux limits are indicated.}
\label{ghasinger-F1:Fig7}
\end{figure}

A similar conclusion about a decline of the X--ray selected AGN space 
density at high redshifts can be obtained from the absence of QSOs with 
$z>5$ in all X--ray survey samples so far. (There was a recent announcement 
of a QSO at z=5.2 in the Chandra observation of the HDF--N, but this 
does not change the conclusions discussed below). Figure \ref{ghasinger-F1:Fig7} shows a 
prediction of number counts for high--redshift QSO from
\cite*{ghasinger-F1:Hai99}, according to which a large number of $z>5$ AGN should be 
detected in any deep survey with Chandra. This theoretical model assumes 
the X--ray luminosity function at z=3.5 determined from the ROSAT surveys
and extrapolates it backwards in time assuming a simple hierarchical
CDM model and a constant black hole mass fraction throughout.
The figure also shows limits for the number counts of $z>5$ AGN from
X--ray surveys at varying flux limits. The most distant QSO among $\sim 2000$
objects in the ROSAT Bright Survey (RBS, \cite{ghasinger-F1:Schw00}) has a redshift of 2.8, the 
lack of higher redshift objects is, however, not constraining given the high flux 
limit of this survey. The lack of $z>5$ AGN in the ROSAT Deep and Ultradeep
Surveys (\cite{ghasinger-F1:Schm98,ghasinger-F1:Leh00,ghasinger-F1:Leh01}) is still just consistent with the Haiman \&
Loeb predictions, the highest--redshift object in the UDS is 
RX~J105225.9+571905 at 
z=4.45 (\cite{ghasinger-F1:Schn98}). The Chandra Deep survey, while only about 60\% 
spectroscopically identified, still provides an upper limit for the number counts of 
$z>5$ AGN significantly lower than the prediction, using the 
conservative assumption that less than half of the unidentified objects are at
redshifts larger than 5. Finally, the 400 ksec Chandra observation in the
Hubble Deep Field proper, providing 100\% identifications for 12 sources 
in the field and their highest redshift object at z=4.42 just outside 
the HDF--N also gives an upper limit about a 
factor of three lower than the Haiman \& Loeb prediction.

The information about the space density of X--ray selected AGN is still 
limited by the small number statistics in the deep X--ray surveys which
cover too small a solid angle. More and wider fields have been surveyed 
by both Chandra and XMM--Newton. As soon as the tedious and time consuming 
optical follow--up work in these fields is completed, we will be able 
to learn more about the decline of the X--ray AGN and therefore their 
formation at early redshifts. The possible discrepancy between a declining 
space density of optical and radio--selected QSOs above a redshift of 2.7
and an apparently constant space density of X--ray selected AGN with a 
decline beyond a redshift of $\sim 4$ could still be understood in terms
of the different luminosity and therefore different black hole mass of
the objects involved. The optical and radio surveys cover a large solid 
angle to a modest flux limit and therefore pick up only the most luminous 
and therefore most massive objects at high redshift. The deep pencil 
beam surveys, on the other hand, sample a much smaller volume to much
fainter flux limits and therefore select high--redshift AGN which are 
intrinsically a factor of more than 10 less luminous and therefore 
probably less massive than the objects selected in wide--angle surveys.
In the hierarchical large scale structure formation the smaller cold dark 
matter halos collapse earlier than the larger ones. Given the correlation between 
black hole mass and galaxy mass (and presumably dark matter mass), it 
is expected that the lower mass black holes are formed earlier than the 
most massive objects and thus that lower luminosity AGN appear earlier
than the most luminous QSOs (see \cite{ghasinger-F1:Kau00}). 
This concept can be tested with more 
optical identifications of Chandra and XMM--Newton surveys and with 
future, even more sensitive X--ray telescopes, like the ESA/ISAS XEUS
mission.

\begin{acknowledgements}

I thank the organisers of the conference ''New Visions of the X--ray Universe''
for the invitation for this review. I thank my co--workers in  the Chandra 
Deep Field South and Lockman Hole 
identification teams for the fruitful collaboration and the permission
to show data in advance of publication. 
This work was supported by DLR grant 55 OR 9908.

\end{acknowledgements}

\end{document}